\documentclass[twocolumn,showpacs,preprintnumbers,amsmath,amssymb]{revtex4}
  \usepackage{graphicx}
  \usepackage{dcolumn}
  \usepackage{bm}
 
\begin{document}

\title{Different doping from apical and planar oxygen vacancies in 
Ba$_{2}$CuO$_{4-\delta}$ and La$_{2}$CuO$_{4-\delta}$}

  \author{T. Jarlborg$^1$, B. Barbiellini$^2$, 
          R.S. Markiewicz$^2$, A. Bansil$^2$.}

  \affiliation{
  $^1$DPMC, University of Geneva, 24 Quai Ernest-Ansermet, 
  CH-1211 Geneva 4, Switzerland
  \\
  $^2$Department of Physics, Northeastern University, Boston, 
  Massachusetts 02115, USA
}
 
 
\begin{abstract}
First principles band-structure calculations for large supercells 
of Ba$_{2}$CuO$_{4-\delta}$ and La$_{2}$CuO$_{4-\delta}$
with different distributions and concentrations of oxygen vacancies show that the effective doping 
on copper sites strongly depends on where the vacancy is located. 
A vacancy within the Cu layer produces a weak doping effect while a vacancy located at an 
apical oxygen site acts as a stronger electron dopant on the copper layers and gradually
brings the electronic structure close to that of La$_{2-x}$Sr$_x$CuO$_{4}$. These effects are 
robust and only depend marginally on lattice distortions. 
Our results show that deoxygenation can reduce the effect of traditional La/Sr or La/Nd substitutions. 
Our study clearly identifies location of the dopant in the crystal structure as an important factor 
in doping of the cuprate planes. 
  \end{abstract}
 
 
  \maketitle
 
\section{Introduction}
Recent work on high-$T_c$ cuprates has explored the role of 
oxygen defects and ordering of oxygen into superstructures. For instance, 
x-ray diffraction measurements on small patches of copper oxide
superconductors reveal structures which may promote high temperature
superconductivity \cite{frat}. 
Other studies involving O vacancies in  
the high-$T_c$ cuprate Sr$_2$CuO$_{4-\delta}$ (SCO) 
have shown that the 
superconducting $T_c$ may become surprisingly large, reaching $95$ K for certain
heat treatments \cite{geba,gao}.
These superconductors have the same 214 structure as La$_2$CuO$_4$ (LCO) 
and it is believed that the La substitution with Sr is associated with
vacancy formation on the oxygen lattice. Interestingly, up to 30 percent of the oxygen might 
be missing in the superconducting samples \cite{geba}. 
This finding is at odds with the doping dependence of $T_c$ in 
La$_{2-x}$Sr$_x$CuO$_4$ (LSCO), since the maximum $T_c$ is found for a moderate hole doping
$x$ of about 0.15 holes per Cu.
It is clear thus that it is important to understand the role 
of oxygen vacancies in doping the Cu-O planes, 
which are widely believed to be the seat of superconductivity. 
So motivated, we examine in this article how location of the oxygen 
vacancies can affect the doping of cuprate planes \cite{raveu,liebau}.
An outline of this paper is as follows. In Sec. II, we present 
the  details of the electronic structure computations in large
supercells. The theoretical results are presented and discussed 
in Sec. III, and the conclusions are summarized in Sec. IV.

\section{Method of calculations}
We specifically extract position dependent mechanisms 
by which  the oxygen vacancies in Ba$_2$CuO$_4$ (BCO) and La$_2$CuO$_4$
modify the electronic structure.
Since Ba is isoelectronic with Sr,
our results for the valence electrons are also applicable
to Sr$_{2}$CuO$_{4-\delta}$.\cite{AB1,bbtj}
Self-consistent first principles calculations are performed 
for (4,2,2) extensions of the basic cell
 \cite{basic} into supercells containing 16 formula units 
with 112 atomic sites in total, as well as for smaller cells
with 7 and 14 sites, by
using the Linear Muffin-Tin Orbital 
(LMTO) method \cite{lmto} 
and the Local Density Approximation (LDA) \cite{lda}. 
While the LDA does not provide a satisfactory description of 
the electronic structure
in underdoped cuprates because of 
inadequate treatment of correlations \cite{ldau}, 
good agreement is found between LDA and various experiments 
\cite{ARPES,XAS,IXS} for optimally and overdoped cuprates. 
The present calculations consider mostly the metallic regime 
where LDA is expected to be a reasonable approximation.
The converged self-consistent results are obtained
using a mesh of 125 k-points
within the irreducible Brillouin zone.
A precise tetrahedron method is used to determine
the density-of-states (DOS) \cite{rath}.   
All sites in the cell are considered as non-equivalent 
throughout the self-consistency cycle. 
We have previously used supercell computations along these lines to search for weak ferromagnetism around Ba-clusters
in La$_{2-x}$Ba$_x$CuO$_4$ (LBCO) \cite{bbtj}. 
However, the present calculations are mostly non spin-polarized,
since magnetism is not an issue here. 
In fact, earlier self-consistent spin-polarized band-structure 
calculations with 28-atom unit cells
have revealed that oxygen vacancies strongly reduce the tendency toward 
antiferromagnetism in LCO \cite{sterne}.
The LMTO results for nonmagnetic La$_2$CuO$_4$ 
are in excellent agreement with full 
potential calculations \cite{freeman} 
demonstrating the high quality of our basis set. 

Vacancies (V) on oxygen sites are introduced in such a way 
as to avoid clustering of vacancies. 
We consider configurations with number of vacancies $n_V$ 
equal to 1, 5, 8, 9, 10 or 11.
The corresponding oxygen deficiencies for Ba$_2$CuO$_{4-\delta}$
are given by $\delta=n_V/16$. Thus 
$\delta$ ranges between 0.06 and 0.69.
The maximum $T_c$ in the SCO system is reported to occur 
for $\delta \approx 0.6$, and another dome of high $T_c$ 
in an uninvestigated region of the phase diagram has been proposed 
\cite{geba}. 
The same type of supercells with $n_V$ equal to 5 and 9 are used 
in calculations with La replacing Ba in order to compare 
the trends for Cu-$d$ band filling in BCO and LCO. 
The lattice constant is kept fixed
in all calculations. 
The present computations are performed with the oxygen vacancies 
either in the planes or in apical sites.

\section{Results and discussion}

\begin{figure}[h]
  \begin{center}
  \includegraphics[width=3.5in]{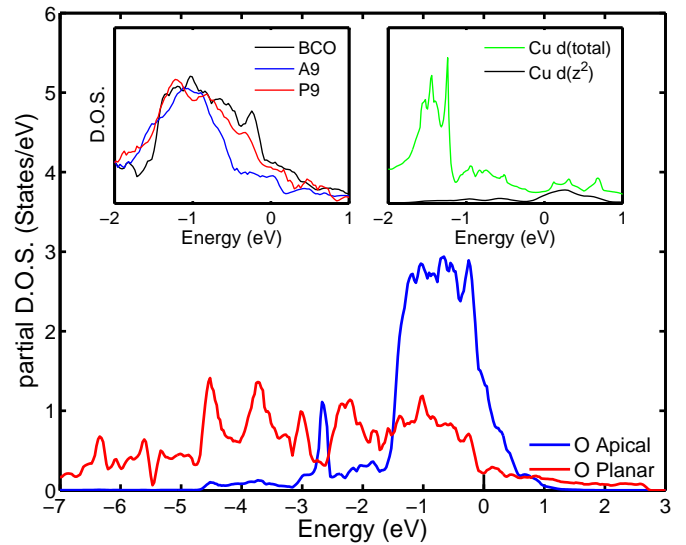}
  \end{center}
  \caption{(Color online) 
  Local oxygen-$p$ part of 
  the DOS on one
  apical and one planar oxygen site in 
  pristine Ba$_{32}$Cu$_{16}$O$_{64}$.
  Left insert: DOS near the Fermi energy 
  for 3 configurations, Ba$_{32}$Cu$_{16}$O$_{64}$ (BCO)
  and vacancy doped Ba$_{32}$Cu$_{16}$O$_{55}$V$_9$, 
  with the 9 oxygen vacancies in
  plane (P9) and apical (A9) positions, respectively, 
  all calculated from 125 k-points.
  Right insert: Local copper-$d$ part of the DOS (total and $d(z^2)$  
  contributions) near the Fermi energy.
  All the DOS plots include a small broadening of 5 meV.} 
  \label{fig1}
  \end{figure}

The Cu-$d$ band filling can be monitored 
through inspection of the DOS or the effective charge within the Cu muffin-tin spheres.
The DOS at the Fermi energy, $N(E_F)$, is dominated by Cu-$d$ 
electrons in all high-$T_c$
cuprates, and the mechanism of superconductivity likely involves the Cu-$3d$ and O-$2p$
character on the Fermi surface (FS). Therefore, it is important to see to what extent the $d$-band filling is affected by
oxygen vacancies or other dopants.
In the left insert to Fig.~\ref{fig1} is shown the total DOS for  
undoped BCO (for the same supercell as for doped cases), 
and with 9 vacancies either in the planes or at the apical positions. Despite the unavoidable noise in the DOS, there is a clear 
d-band shift to lower energy (with 
respect the Fermi level) for vacancies on apical sites   
compared to the pure case (BCO), while the shift for vacancies 
in the planes  
is much smaller with even some small enhancement of the DOS above $E_F$.
A DOS enhancement produced by O vacancies
was found in coherent-potential-approximation (CPA)
calculations \cite{cpa_papa,deweert} where 
all the oxygen sites (planar and apical)
had equivalent random occupation. Notably, rigid band like pictures 
\cite{bansil1,bansil2} 
have often been invoked in describing the doping evolution of 
the overdoped and optimally doped cuprates. 
Our large supercell treatment here goes beyond 
the simple rigid band models or the CPA-type mean-field approaches
\cite{bansil3} to elucidate the local electronic and magnetic 
properties of the system. 

Insight into the underlying physics is provided by the main 
frame of Fig.~\ref{fig1}, which compares the
partial DOS of planar and apical oxygen in the undoped parent compound.  
Note that since Ba donates one less electron than La, the Fermi level 
is shifted downward compared to the situation in LSCO.  In this case, 
bands of both $d_{x^2-y^2}$ and $d_{z^2}$ character are present
near the Fermi level, with the latter dominating the DOS 
as shown by right hand side insert in Fig.~\ref{fig1}.
Consequently, the partial DOS of the apical O 
near the Fermi level is larger, 
since the electrons of these O atoms 
hybridize with the Cu $d_{z^2}$ electrons.

\begin{figure}[h]
  \begin{center}
  \includegraphics[width=3.5in]{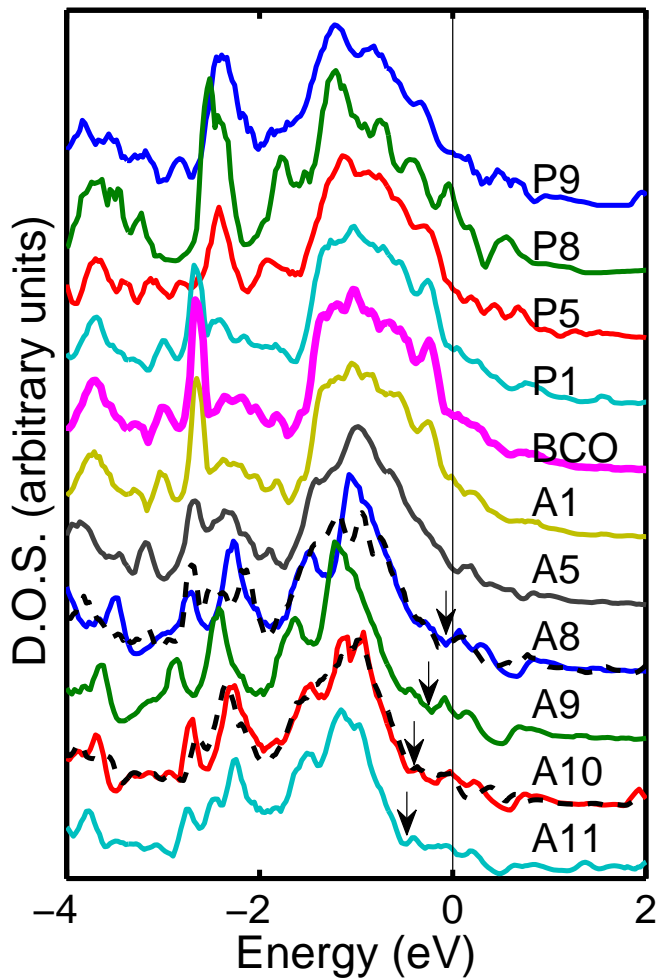}
  \end{center}
  \caption{(Color online)
  DOS near the Fermi energy for various vacancy configurations in Ba$_{32}$Cu$_{16}$O$_{64-n}$.
  BCO is the undoped case, ``Pn" refers to $n$ vacancies in the plane and ``An" refers to $n$ apical vacancies.
  The DOS plots include a small broadening of 5 meV. The Fermi energy is taken as zero in all cases.
  Black dashed lines show the DOS for different arrangements of the same number and type of O vacancies. 
This illustrates the robustness of the main trends.
When $n\ge 8$ for the apical vacancies, the arrows indicate the onset of the sharp rise associated with the $d_{z^2}$ 
band below the Fermi level.}
  \label{fig2}
  \end{figure}

Figure~\ref{fig2} shows the total DOS for
pristine BCO (for the same supercell as for doped cases),
and with up to 9 vacancies in the planes or up to 11 vacancies in the apical positions.  
Apical site deoxygenation shows a clear trend, 
shifting the $d$-band to lower energy.  The effect is also consistent with 
the change of the number of valence electrons within the Cu muffin-tin
spheres for different vacancy configurations, shown 
in Table~\ref{table1}. The strikingly different DOS shapes of apical and planar O-p bands 
in BCO suggest that x-ray emission spectroscopy 
(XES) would be a good experimental method for detection of doping dependencies. Furthermore, the $d$ band 
shift below $E_F$ seen in Fig.~\ref{fig2} for the apical but not for planar 
O vacancies would be a key signature to look for with XPS.
Notably, recent O K-edge x-ray absorption (XAS) experiments 
on SCO/LCO superlattices \cite{smadici} have shown that the reduction 
of the second hole peak in the unoccupied DOS of SCO provides a signature of 
apical oxygen vacancies in SCO. In fact, earlier XAS 
measurements on overdoped LSCO \cite{chen} and electron energy-loss spectroscopy 
on SCO \cite{yang} have determined that 
such peaks belong to the apical oxygen unoccupied partial DOS. 
Clearly, changes in these unoccupied DOS peaks \cite{XAS} mirror 
trends shown in Fig.~\ref{fig2} associated with the apical 
oxygen character in the occupied DOS.

\begin{table}[ht]
\caption{\label{table1}
 Average valence charge of Cu ($Q_{Cu}$, within the atom sphere radius of 0.335$a_0$) and its difference
 relative to the case without vacancy ($\Delta Q_{Cu}$) for different number of oxygen vacancies $n_V$
 within planar and apical positions
 in Ba$_{32}$Cu$_{16}$O$_{64-n_V}$V$_{n_V}$ (BCO) and La$_{32}$Cu$_{16}$O$_{64-n_V}$V$_{n_V}$ (LCO).
 The corresponding average valence charge of the oxygen vacancy ($Q_{V}$ within atomic sphere radii
 of 0.317$a_0$ and 0.345$a_0$ for planar and apical sites, respectively) is also reported.
 Since there are in total 64 oxygen sites
 in the $112$ atom cell, the effective doping is $\delta = n_V/16$. Different distributions of $8$
 and $10$ vacancies on apical sites for BCO show very little difference in the charge
 transfer. The cases $n_V=1$ with or without lattice distortion are almost identical, see text.}
  \vskip 2mm
  \begin{center}
  \begin{tabular}{l c c c c c c}
  \hline
     & BCO  & BCO  & BCO  & LCO & LCO& LCO\\
   $n_V$ &$Q_{Cu}$ & $\Delta Q_{Cu}$ &$Q_{V}$ & $Q_{Cu}$ &  $\Delta Q_{Cu}$ &$Q_{V}$\\
  \hline \hline
  9 (plane)  & 10.054 & -0.010 &0.662 & 10.331 & -0.063 & 0.889\\
  8 (plane)  & 10.055 & -0.009 &0.655 &  - & -& - \\
  5 (plane)  & 10.057 & -0.007 &0.650 & 10.368 & -0.026& 0.877 \\
  1 (plane)  & 10.062 & -0.002 &0.640 & - & -& - \\
  0 & 10.064 &  0.000 & - & 10.394   & 0.000& - \\
  1 (apical) & 10.071 &  0.007 &0.303 & - & - & -\\
  5 (apical) & 10.107 &  0.043 &0.313 & 10.450 & 0.056& 0.623 \\
  8 (apical) & 10.142 &  0.078 &0.320 & - & - & -\\
  8 (apical) & 10.145 &  0.081 &0.308 & - & - \\
  9 (apical) & 10.164 &  0.100 &0.327 & 10.456 & 0.062& 0.662\\
 10 (apical) & 10.179 &  0.115 &0.326 & - & - & -\\
 10 (apical) & 10.180 &  0.116 &0.329 & - & - & -\\
 11 (apical) & 10.200 &  0.137 &0.336 & - & - & -\\
  \hline
  \end{tabular}
  \end{center}
  \end{table}
The Cu muffin-tin spheres have the same radius in all cases. 
For an increased number of apical vacancies in BCO there is 
a monotonic increase of the Cu charge so that about 0.1 electrons are
added to the sphere for $\delta \approx 0.56$. The fine details of the vacancy distribution 
do not seem to be important since results from two different configurations 
(for 8 and 10 vacancies) give almost identical results. 
In BCO with vacancies within the CuO$_2$ planes, 
there is little modification of the 
Cu charge, about 1/10 of the change due to apical vacancies. 
These small changes shown in Table \ref{table1} indicate less Cu charge
for  O vacancies within the CuO$_2$ plane, i.e. there is an effective hole doping within the Cu band.

The average valence charge of the oxygen vacancy $Q_V$ in Table I 
yields a simple physical
picture of the difference between vacancies in planar and apical O-positions.
When the oxygen vacancy is located in the Cu-O planes, it acts as an electron trap (i.e. $Q_V$ is large),
while when the vacancy is in the charge reservoir, the electrons are more attracted by the Cu-O planes than by the vacancy.
Hence, apical vacancies behave as expected from simple valence arguments: 
O$^{2-}\rightarrow V + 2e$, which dopes the 
planes with two electrons per vacancy.  
On the other hand, planar vacancies are anomalously charged, so : 
O$^{2-}\rightarrow V^- + e$, 
resulting in a weaker electron doping.  
Note that in both cases, electrons are doped into the planes, but for planar O vacancies, they are more localized on the vacancy.
Apical oxygen vacancies will thus mainly remove O-p bands in
the interval where the apical O-DOS is high ranging from $E_F$ down to 
1.5 eV below $E_F$.
In contrast, the DOS reduction due to planar oxygen vacancies will 
be spread over a wider energy range because of the wider
partial DOS for planar oxygen sites shown by Fig.~\ref{fig1}.

We also carried out computations on LCO to further demonstrate the robustness of our results. 
In particular as illustrated in Table \ref{table1}, the electron doping from apical O-vacancies in 
LCO is comparable to that of BCO. Moreover, 
to estimate the effect of lattice relaxation around vacancies, calculations were made with
a vacancy on one apical oxygen position for BCO, where a large lattice relaxation is imposed. 
The two atoms surrounding the missing O site, namely the Ba above (along the $c$ axis) and the Cu below 
were allowed to fill in the empty space by moving 0.03$a_0$ towards the center of the missing O. 
The amplitude of these distortions
is rather large, corresponding to a thermal distortion amplitude 
for a temperature of several hundred degrees Kelvin \cite{tj7}. 
This distortion is found to make very little change in the number of electrons on the Cu sites. 
No Cu atom changes its charge more than 0.005 electrons. However, the Cu atom
next to the vacancy gains 0.05 electrons compared to Cu in ideal BCO, and this is independent of
relaxation. The Ba atom looses 0.08 electrons
and the vacancy (with the same volume as the O atom) gains 
0.10 electrons. Other sites typically change their charge
by $\pm$0.01 electrons or less. Therefore, our finding of 
net charge transfer due to high vacancy concentration
is essentially independent of lattice relaxations.

\begin{figure}[h]
  \begin{center}
  \includegraphics[width=8.5cm]{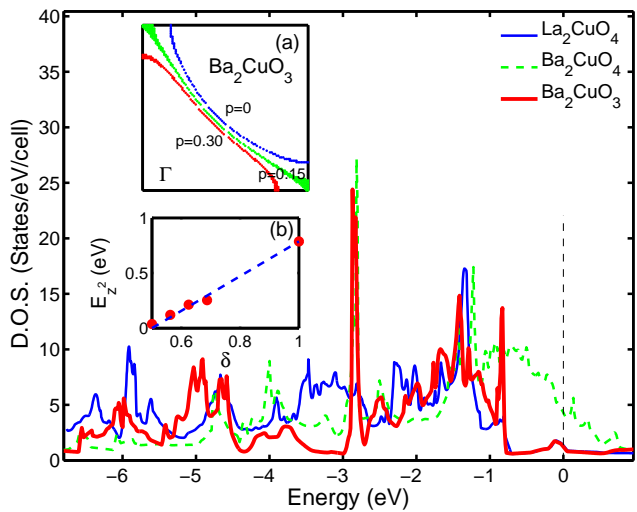}
  \end{center}
  \caption{(Color online)
  DOS near the Fermi energy for Ba$_2$CuO$_3$, Ba$_2$CuO$_4$ 
and La$_2$CuO$_4$. 
  Insert (a) shows the evolution of the Fermi surface 
in the $k_z = 0$-plane in Ba$_2$CuO$_3$
  as a function of the rigid-band doping for 0, 0.15 and 
0.30 holes per unit cell, while insert 
(b) shows how the Fermi level shifts above the $d_{z^2}$ band onset as vacancies are added.}
  \label{newfig}
  \end{figure}

To better understand the role of apical vacancies in doping the planes, 
we also present 
some calculations for a related compound with even more apical O vacancies:
Ba$_2$CuO$_3$ (BCO-3) where one layer of apicals is missing 
\cite{footnote1}. 
Figure~\ref{newfig} compares the DOS of BCO-3 to stoichiometric Ba$_2$CuO$_4$ (BCO-4) and La$_2$CuO$_4$ (LCO-4), 
demonstrating that the DOS of BCO-3 and LCO-4 are almost identical near $E_F$ -- i.e., 
for the $d_{x^2-y^2}$ band. At lower energies, from 1 eV below $E_F$, 
there are differences because of the removal of three hybridized O-p
bands. In contrast, the DOS of BCO-4 is very different even near $E_F$. 
These DOS functions are consistent with the trend shown in Fig.~\ref{fig2}, despite the 
different compositions and orderings. Disorder has some effect on the details but not on the main features of the DOS.  
The Cu valence charge in BCO-3, 10.25 electrons, is consistent with the results in
Table \ref{table1} for an extrapolated concentration of 16 apical vacancies. 
Thus, the DOS and FS of the relevant Cu-O single band of BCO-3 and LCO-4
behave in the same way, but for an effective Cu charge 
that is slightly smaller in BCO-3: about 10.25 vs. 10.39 in LCO-4, see Table \ref{table1}.  
There is a striking similarity of the FS of BCO-3 (shown in the upper inset of Fig.~\ref{newfig}) and the well-known
FS of LCO-4 (see ref. \cite{FSpaper}). Furthermore, it can be verified that
the FS's of the two systems have an almost identical evolution with the number of holes $p$ 
in the single band at $E_F$ as shown 
in the insert of Fig.~\ref{newfig} for BCO-3.

We can now understand that the role of apical vacancy doping is 
to push the Fermi level from the middle of the $d_{z^2}$ band 
for $n_V=0$ 
to the middle of the $d_{x^2-y^2}$ band for $n_V=16$.  
For $n_V\ge 8$, the sharp rise associated with the $d_{z^2}$ 
band is located below the Fermi level as shown by the arrows 
in Fig.~\ref{fig2}.  
We plot the energy shift between the $d_{z^2}$ onset and $E_F$ as a function of 
$\delta=n_V/16$ in insert (b) of Fig.~\ref{newfig}.  
Remarkably, the shift for BCO-3 also fits on the same line.  
A linear shift is expected if the DOS is 
approximately constant, as in Fig.~\ref{newfig}.  Hence, we find that in BCO for 
$\delta$ between 0.5 and 1, the vacancies dope electrons into 
an essentially empty $d_{x^2-y^2}$ band, and the physics 
should be similar to that of strongly hole-overdoped LCO.

In comparing our results to experiments on the Sr$_2$CuO$_{4-\delta}$ system, 
we note that  Gao {\em et al.} \cite{gao}
conclude that the missing oxygens are outside the CuO plane and that the structure remains
stable even at this high doping \cite{jin}. 
These conclusion are also corroborated by our calculated total energies for having 1 apical
or 1 planar vacancy in the undistorted 112 site supercell, 
since the total energy 
is considerably lower
for the apical position, 
of the order of 1.5 eV.
Vacancy ordering leading to long range periodicity is
suspected to be important for boosting $T_c$ to high values 
\cite{chma,frat,apl,mare}. 
Orderings of apicals or other defects, leading to long-range periodicity,
are expected to induce a lowering of the total energy because of the appearance of a partial
gap at $E_F$ \cite{tj7}.

Since each apical vacancy contributes two electrons to the CuO$_2$ plane, 
the addition of apical vacancies could lead 
to a new route for electron-doping LCO.  We note that 
O vacancies in the reservoir layers also provide electron doping in  
YBa$_2$Cu$_3$O$_{7-y}$ (YBCO) \cite{ybco}, even though YBCO is structurally 
very different from both LCO and BCO.
The doping for high-$T_c$s in BCO cuprates corresponds to a nearly empty 
$d_{x^2-y^2}$ band, 
consistent with the suggestion of a second superconducting dome.  
It is interesting to note that dopings for high-$T_c$ in pnictides 
often seem to be associated with the Van Hove singularity 
at a band onset \cite{Borisenko}, although the reason for this is not clear. 
In the case of BCO, the possibility of 
altered $d_{z^2}$-$d_{x^2-y^2}$-orbital character near $E_F$ 
due to vacancy ordering could also modify the superconducting properties. 
  
\section{Conclusion} 
Our study shows that properties of 
copper oxides high temperature superconductors 
depend dramatically on the concentration and position of oxygen vacancies. Our results indicate that manipulation of apical oxygen vacancies is an effective pathway for optimizing the electronic structure for higher $T_c$.
The similarity between the band structure for apical vacancy doped BCO and the commonly doped LCO system is striking, but consistent with 
the notion that the high $T_c$ in Sr$_2$CuO$_{4-\delta}$ is 
due to another "dome" of superconductivity with 
a different hole doping on Cu-sites. The slight difference in Cu valence charge
between the two systems can affect the magnetic response despite 
a very similar Fermi surface. Another implication of our study 
is that O vacancy formation can compensate
the doping resulting through conventional La substitutions.\\
{\bf Acknowledgements}\\
We acknowledge useful discussions with M. Marezio, C. Q. Jin and A. 
Bianconi. This work is supported by the U.S.D.O.E. contract DE-FG02-07ER46352. It benefited from the allocation of supercomputer time at NERSC and Northeastern University's Advanced Scientific Computation Center (ASCC), theory support at the Advanced Light Source (grant number DE-AC02-05CH11231), and the Computational Materials and Chemical Sciences Network (CMCSN) program of the Division of Materials Science and Engineering, U.S.D.O.E, under grant number DE-SC0007091.


\begin{thebibliography}{10}
\bibitem{AB1} 
The advantage with BCO compared to SCO is that the simple virtual crystal approximation scheme can be used to capture aspects of the electronic structure changes induced by Ba/La substitutions as an alternative to the more complicated supercell calculations \cite{bbtj}.
\bibitem{basic} Atomic positions (in units of the lattice constant $a_0$=3.787 \AA) within
 the basic cell are $(0,0,\pm1.265)$ for Ba/La, $(\frac{1}{2},0,0)$ and $(0,\frac{1}{2},0)$ for planar O,
 $(0,0,\pm0.641)$ for apicals and $(0,0,0)$ for Cu. The ratio $c/a_0$ is 1.754.
\bibitem{footnote1}
Calculations for three different configurations of two apical vacancies in the double cell (Ba$_4$Cu$_2$O$_6$,
assuming the same unit cell volume and no distortions) show that the configuration with one layer of missing oxygens has
a lower total energy than the other two possible configurations. The differences are 1 to 2 eV per cell
with the least stable
configuration when two apical vacancies surround one of the Cu atoms. The DOS for 
the checkerboard configuration of missing O above or below Cu is not too different from the DOS shown
in Fig.~\ref{newfig}, while there is a pseudogap at $E_F$ if the two vacancies surround both sides of one Cu.
Spin polarized calculations for the double cells (Ba$_4$Cu$_2$O$_6$ and La$_4$Cu$_2$O$_8$)
with imposed anti-ferromagnetic order on Cu give a smaller exchange enhancement in the former case by factors of
0.6 to 0.8 relative to LCO, depending on how the two apical vacancies are distributed.
\end{thebibliography}
  \end{document}